\documentclass[12pt,preprint]{aastex}
\usepackage{natbib}
\slugcomment{To appear in {\it The Astrophysical Journal}}

\shorttitle{branch}
\shortauthors{Murray and Lin}

\begin{document}
\title{Energy Dissipation in Multi-Phase Infalling Clouds in Galaxy Halos}
\author{Stephen D. Murray}
\affil{Lawrence Livermore National Laboratory, L-22, P.O. Box 808,
Livermore, CA, 94550\\Electronic mail: sdmurray@llnl.gov}

\and

\author{Douglas N. C. Lin} \affil{Department of Astronomy \&
Astrophysics, University of California, Santa Cruz CA 95064}

\begin{abstract}
During the epoch of large galaxy formation, thermal instability leads
to the formation of a population of cool fragments which are embedded
within a background of tenuous hot gas. The hot gas attains a quasi-hydrostatic
equilibrium. Although the cool clouds are pressure confined by the hot gas,
they fall into the galactic potential, and their motion is
subject to drag from the hot gas.  The release of gravitational energy
due to the infall of the cool clouds is first converted into their
kinetic energy, and is subsequently dissipated as heat.  The cool
clouds therefore represent a potentially significant energy source
for the background hot gas, depending upon the ratio of thermal energy
deposited within the clouds versus the hot gas.  In this
paper, we show that most of dissipated energy is deposited in the
tenuous hot halo gas, providing a source of internal energy to
replenish losses in the hot gas through bremsstrahlung emission and
conduction into the cool clouds.  The heating from the motion of the cool
clouds allows the multi-phase structure of the interstellar medium to
be maintained.

\end{abstract}

\section{Introduction}
The stellar velocity dispersion in the halos of galaxies similar to the
Milky Way exceeds 100~km~s$^{-1}$.  The gravitational potential that binds
the stars to their host galaxies is dominated by collisionless dark matter.
According to the widely adopted cold dark matter (CDM) scenario, these
normal galaxies are formed through the mergers of much smaller
entities, dwarf galaxies.  After violent relaxation, the dark matter
is well mixed in phase space and attains an extended 3-D spatial
distribution.  In spiral galaxies, the formation and concentration of
stars in extended, flattened, rotating disks requires the detachment
of ordinary matter from the dark-matter halos of the original host
dwarf galaxies.  The dominance of the dark-matter halo to the galactic
potential at large radii, and the separation of the ordinary matter
imply that, during the epoch of galactic buildup, the ordinary matter was
primarily in the form of gas which dissipated a substantial fraction
of its initial potential energy.

In a previous paper (Lin \& Murray 2000), we considered the dynamical
evolution of infalling gas in the halos of normal galaxies.  We showed
that for typical values ($\sim 10^6$ K) of the virial temperature, the
cooling timescale increases with temperature, and the protogalactic
clouds (hereafter PGC's) are thermally unstable (Field 1965). Thermal
instability leads to the rapid growth of perturbations and
fragmentation of a PGC (Murray \& Lin 1990). The result is that a
two-phase medium develops during the initial cooling of the PGC, in
which a population of warm fragmentary clouds (WFC's) are confined by
the pressure of hot, residual halo gas (RHG) (Burkert \& Lin 2001).
The RHG is cooled by radiative emission and conductive transport into the
WFC's (which are efficient radiators).  In our earlier work, we assumed
that the RHG is heated primarily by the release of the gravitational
energy as the WFC's into the central region of the halo potential, due 
both to their collective gravity as well as that of the dark matter.
The WFC's are unable to cool
below 10$^4$~K until their density reaches a sufficiently high value
that the WFC's become self-shielded from external photo-dissociating
UV radiation (Couchman \& Rees 1986; Haiman, Rees, \& Loeb 1997; Dong,
Lin, \& Murray 2003).

In the above picture, the evolution of the WFC's is similar to that of
Lyman-$\alpha$ clouds and high velocity clouds (HVC's).  Both of those systems
have been proposed as representatives of late-time accretion of material
in an ongoing process of galaxy buildup by mergers \citep{MI94b,
Blitzetal99,Manning99}.  Because they evolve at an earlier time and closer
to the centers of the parent galaxies, however, the WFC's would evolve in
an environment of higher pressures and UV fluxes, compared to either
Lyman-$\alpha$ clouds or HVC's.  Their environment may, instead, more closely
resemble that of cooling flows (e.g.
Sarazin 1986; Loewenstein \& Mathews 1987; Sarazin \& White 1987, 1988),
and many of our results may have relevance to those systems.  Additionally,
the Ly-$\alpha$ clouds have been proposed as being contained within
dark matter ``minihaloes,'' (e.g. Rees 1986; Ikeuchi 1986;
Mo, Miralda-Escude, \& Rees 1993) whereas the WFC's are either pressure
confined, or at most weakly self-gravitating

In this paper, we verify our basic conjecture that most of the
gravitational energy released by the infalling WFC's is dissipated within
the RHG.  That process is crucial to the assumption that the RHG is in
quasi-thermal equilibrium.
Without this heating source, the background gas would gradually be
depleted due to loss of thermal energy and precipitation into WFC's.
A reduction in
the pressure of the background gas would also enable the WFC's to expand
and eventually eliminate the multi-phase structure of the gas.  In
order to simulate this process in detailed, we adopt a 2-D numerical
hydrodynamic scheme with a multi-phase medium.

The motion of clouds relative to, and their interaction with an external
medium has been studied by numerous authors.  \cite{MI94a} examined
ram-pressure stripping due to the supersonic motion of gas past clouds
confined within minihalos, a very different situation from that described
above for the evolution of the WFC's.  Tenorio-Tagle et al. (1986, 1987)
examined the interactions of clouds hitting relatively high density galactic
disks at high speeds.  Again, that is a very different situation from the
evolution of the WFC's, which move slowly through a low density medium
with a smooth density distribution.  \cite{MWBL93} examined the loss of
gas from a cloud due to the growth of Kelvin-Helmholtz instability for
transsonic motions.  As with the above studies, however, the energy transfer
between the cloud and the background gas was not examined.

We proceed by briefly describing
our method and the model parameters in \S2.  In \S3, we analyze the
results of our computations.  Finally we discuss the implication of
these results in \S4.

\section{Numerical Method and Model Parameters}

\subsection{Equation of Motion}

Following its collapse into the potential of the galactic dark matter
halo, the RHG is shock-heated to the virial temperature of the
potential, and rapidly attains a quasi hydrostatic equilibrium.  For
computational simplicity, we adopt a Cartesian coordinate system in
which the galactic potential $g$ is imposed in the $y$ direction.  For
a spherically symmetric potential, $y$ corresponds to the radial
direction.  The equation of motion of the RHG becomes
\begin{equation}
{d V_{h x} \over d x} = - {1 \over \rho_h} {d P_t \over d x}
\end{equation}
\begin{equation}
{d V_{h y} \over d y} = - {1 \over \rho_h} {d P_t \over d y} -g
\end{equation}
where $\rho_h$, $P_h$, $V_{h, x}$, and $V_{h,y}$ are respectively the
density, pressure, two-velocity components of the RHG, $P_t = P_h +
P_w$ is the total pressure, $\rho_w$, $P_w$, $V_{w, x}$, and $V_{w,y}$
are respectively the density, pressure and two velocity components of
WFC's.  The equation of motion for the WFC's is similar,
\begin{equation}
{d V_{w x} \over d x} = - {1 \over \rho_w} {d P_t \over d x}
\end{equation}
\begin{equation}
{d V_{w y} \over d y} = - {1 \over \rho_w} {d P_t \over d y} -g+F_D,
\end{equation}
where $F_D$ is a drag force term, which is a function of the speed and
geometry of the WFC's, and of their density contrast with the RHG.

\subsection{Parameters for the Residual Halo Gas}

We consider four models.  The parameters are listed in
Table~\ref{tab:models}, which lists, for each model, value of $g$, the
polytropic index for the cloud, $\gamma_w$,
the density contrast between the cloud and the 
background, $D_\rho$, and the initial downward speed of the cloud,
normalized to the sound speed of the background.
In all cases, the RHG is initialized
with the same temperature throughout, but thereafter evolves with
a polytropic equation of state, in which $P_h = K_h \rho_h ^{\gamma_h}$
where $K_h$ is the adiabatic constant, and the polytropic index
$\gamma_h=5/3$ for each model.  We also assume that the RHG is initially
in hydrostatic equilibrium, such that
\begin{equation}
\rho_h=\rho_0{\rm e}^{{g\over{C_h^2}}\left(y-y_0\right)},
\label{eq:den}
\end{equation} 
where $\rho_0$ is the density at a reference height $y_0$, and $C_h$ is the
isothermal sound speed of the RHG.  Because the RHG initially has the
same temperature throughout, the magnitude of $K_h$ is a function of 
$y$, such that
\begin{equation}
K_h = C_h^2 \rho(y)^{1-\gamma_h}.
\end{equation}
The density scale height of the RHG is 
\begin{equation}
r_h = { C_h^2 \over g}.
\end{equation}

In all models the velocities are normalized to $C_h=1$.  The initial
location of the WFC's is set to be at $x_0=y_0=0$.  The value of $g$
is uniform throughout the grid, justified by the fact that the 
computational domain represents a small fraction of a galaxy.  For
Models~1-3, we set $g=0.1$, so that $r_h = 10$, while in Model 4,
$g=0.05$, giving $r_h = 20$.  We set $\rho_0=1$.  In these units,
Equation~(\ref{eq:den}) reduces to
\begin{equation} 
\rho_h = {\rm e}^{-g\left(y - y_0\right)}.
\end{equation}
The computational domain extends from -0.75 to 0.75 in $x$, and from
-15 to 1 in $y$. At the base of the computational domain, $\rho_h
/\rho_0 \sim 4.5$ for Models~1-3, and 2.1 for Model 4.

\subsection{Parameters for the Warm Fragmentary Clouds}

The density ratio at the
launch point, $D_{\rho}\equiv \rho_w/\rho_h =10^{2}$ in Models 1-3, while
$D_\rho = 25 $ in Model 4.  The magnitude of $D_\rho$ would be constant 
throughout the simulation time if
1) the WFC's retain their integrity, 2) $\gamma_h = \gamma_w$, and 3)
there is no shock dissipation to modify $K_h$ and $K_w$.  In general,
however, $D_\rho$ is a function of $y$, depending on the equation of
state for both the WFC's and RHG.

The value $D_{\rho}=100$ is selected to represent ionized clouds at
temperatures of 10$^4$~K in
a Milky Way-sized galaxy, for which the RHG is heated to the virial temperature
of $\approx$10$^6$~K.  The smaller value of $D_{\rho}$ for Model~4 would be
appropriate for either cooler backgrounds or warmer clouds.

The physical dimensions of the WFC's are set by dynamical and thermal
processes \citep{LM00}.  Clouds
below a minimum radius, $S_{min}$, are re-heated by conduction from the RHG.
The maximum radius, $S_{max}$, is set by the point at which the clouds
become self-gravitating.  Such clouds have negative specific heats, and so 
are unstable to external heating.  The lower limit upon the cloud size
translates to \citep{LM00}
\begin{equation}
S_{min,s}=1.6{\ T}^{3/2}_6{\ D}_{100}^{-1}{\ n}_{10}^{-1}
{\ }\Lambda_{25}^{-1}{\ \rm pc},
\end{equation}
in the limit of saturated conduction, where $T_6$ in the temperature of the
RHG in units of 10$^6$~K, $D_{100}$ is the density contrast between the
cloud and RHG in units of 100, n$_{10}$ is number density of the cloud in
units of 10~cm$^{-3}$, and $\Lambda_{25}$ is the cooling efficiency in units
of 10$^{-25}$~ergs~cm$^3$~s$^{-1}$, characteristic of low metallicity gas
near 10$^4$~K \citep{DM72}.  In the limit of unsaturated conduction,
\begin{equation}
S_{min,u}=4{\rm T}^{7/4}_6{ \rm n}_{10}^{-1}
{\ }\Lambda_{25}^{-1/2}{\ \rm pc}.
\end{equation}
The maximum cloud size is set by the Bonner-Ebert criterion for self-gravity
to become important,
\begin{equation}
S_{max}=350{\ T}_4\left({{nT}\over{10^5}}\right)^{-1/2}{\ \rm pc},
\end{equation}
where $T_4$ is the temperature of the WFC's in units of 10$^4$~K, and
$nT$ is the pressure of the clouds (assumed to be in pressure equilibrium
with the RHG).

In the two-phase model discussed by \cite{LM00}, the WFC's are heated by
UV emission from massive stars, in a self-regulated star formation process.
For the parameters given above, the total column densities of the clouds
range from $5\times10^{19}$ to $10^{22}$~cm$^{-2}$.

In Models 1-2 and 4, we adopt a polytropic equation of state for the
WFC's with a power index $\gamma_w = \gamma_h$.  In Model~3 we attempt to
maintain $\gamma_w\approx1$ throughout the evolution.  This is done by allowing
the cloud to cool, but turning cooling off below a set minimum temperature,
which we select as $10^4$~K.  The high cooling efficiency in this temperature
regime ensures that the cloud temperature cannot significantly exceed the
cutoff temperature.  Cooling is not allowed to proceed in the background gas.

Because we are using a single-fluid code (see below), zones where cloud and
background gas mix are a concern.  In order to prevent significant cooling of
the background gas, cooling is turned off whenever the background gas exceeds
a volume fraction of 0.5, as measured by the relative amounts of the two
tracers initially placed within the cloud and background.  Cooling is also not
allowed whenever the temperature of
a cell exceeds 0.2 times the initial temperature of the background gas.

The models with polytropic equations of state permit the use of dimensionless
numbers, as used here, to scale the results to a wide range of systems.  The
presence of cooling in Model~3, however, introduces some dimensions into the 
problem.  In that model, we take $T_h=10^6$~K, and $T_w=10^4$~K, appropriate,
as discussed above, for an L$_\ast$ galaxy.  The corresponding sound speeds
are $C_h=130$~km~s$^{-1}$, and $C_w=13$~km~s$^{-1}$ (assuming ionized gas).
The initial density of the cloud, $n_w=6$~cm$^{-3}$.
A length dimension is not explicitly imposed upon the problem, because
the heating and cooling are strictly local processes.  Typical values can,
however, be computed.  For an isothermal potential, $g=V_c^2/R$,
where $V_c$ is the circular speed and $R$ is the galactocentric radius.  Taking
$V_c=220$~km~s$^{-1}$, the physical scale height of the gas
\begin{equation}
R_h={{C_h^2}\over g}={130\over220}R=350~R_{kpc}{\ \rm pc},
\end{equation}
where $R_{kpc}$ is the galactocentric radius in kpc.  In code units,
$R_h=10$ (see previous section), and so one unit of distance in the code
corresponds to $l_{unit}=35$~$R_{kpc}$~pc in physical units.  The cloud radius
is initially set to be 0.2 in code units, or 7~$R_{kpc}$~pc in physical
dimensions.  The unit of time for Model~3 is given by the ratio 
$l_{unit}/C_h=0.26~R_{kpc}$~Myr.

In Models 1, 3, and 4, we assume that the WFC is initially at rest.  The
evolution
of the WFC's is, however, dynamic, with clouds continually colliding and
merging, being disrupted by dynamical instabilities, being reheated by 
conduction from the RHG, cooling to form stars, and condensing out of the
RHG.  When the clouds form from the RHG, they would be expected to have
typical speeds up to the sound speed of the RHG.  In Model 2, therefore,
we take the WFC to be initially falling in the $-y$ direction, at a velocity
equal to $C_h$.

Due to their negative buoyancy, the WFC's fall through the RHG in all
models. If the background RHG is not perturbed, it induces a drag
force $F_D$ on WFC's.  For WFC's with sizes $S$ which are larger than the
mean free path of particles in the RHG,
\begin{equation}
F_D = {1\over2}C_D \pi S^2 \rho_h V_w ^2,
\end{equation}
where $C_D$ is the drag coefficient, and $S$ is the cloud radius
(Batchelor 2000).  In flows with high Reynolds number, the
turbulent wake behind the body provides an effective momentum transfer
mechanism, dominating $C_D$.  For example, the experimentally measured
$C_D$ for a hard sphere in a nearly inviscid fluid is 0.165 (Whipple
1972).  For compressible gas clouds, $C_D$ is probably closer to
unity.

When $F_D\approx g$, the WFC's attain a terminal speed
\begin{equation}
V_t \approx \left( {8 D_{\rho} {S g} \over 3 C_D } \right)^{1/2}.
\label{eq:vterm}
\end{equation}
At the launch point, the size of the WFC is set to be $S (y_0) = 0.2$
in models 1-3. If $C_D =1$ and the WFC preserves its integrity, $V_t =
2.3$, which would exceed sound speed of the RHG. Once the Mach number
of the WFC exceeds unity, however, shock dissipation would greatly increase
the drag relative to the above estimate.  Prior to the WFC achieving 
$V\approx1$, however, Rayleigh Taylor instability causes it to break up
into smaller pieces.  For smaller fragments, the value of
$V_t$ is reduced, as seen in Equation~(\ref{eq:vterm}).  Due to
shock dissipation, the sound speed in the RHG is also slightly larger
than that in Equation~(\ref{eq:vterm}).  Both of the above factors may
prevent the falling WFC's from attaining $V_t>C_h$.
Because the WFC's are pressure confined, however, their internal
sound speed $C_w = D_\rho ^{-1/2} C_h =0.1 \ll V_t$, and so internal
shock dissipation is likely to occur within the WFC's.  In order to
examine the role of the relative magnitudes of the speeds, we choose,
in model 4, $S (y_0) =0.1$, $g=0.05$, and $D_\rho (y_0) = 25$
such that $V_t < C_h$ throughout the computational domain.  Shock
dissipation does not occur in the RHG, but it is present interior to
the WFC.
  
In Model 3, we assume the same initial condition as model 1, but 
adopt an effectively isothermal equation of state for the WFC's by 
allowing the gas to cool to 10$^4$~K.
The resulting energy drainage would lead to a greater
dissipation rate within the WFC's but it should not significantly
modify the energy deposition rate into the RHG.

\subsection{Numerical Method}

The models discussed below are calculated using Cosmos, a multi-dimensional,
chemo-radiation-hydrodynamics code developed at Lawrence Livermore
National Laboratory (Anninos, Fragile \& Murray 2003).  For the current
models, radiative emission is not included.  In order to maximize the
resolution, the models are run in two dimensions.  Because Cosmos runs
on a Cartesian grid, this means that the clouds simulated are actually
slices through infinite cylinders, rather than spheres.  This limitation
should not, however, significantly alter our conclusions, and allows us to
run the simulations at significantly higher resolution than would be
possible in three dimensions.  The resolutions
of the models are 300x3200 zones.  The clouds are therefore resolved by
80 zones across their diameters.  This is somewhat poorer than the
resolution found necessary by Klein, McKee, \& Colella (1994) for their
study of shock-cloud interactions.  Because the clouds in our models are not 
subject to extreme shocks, however, lower resolutions should be adequate,
and reductions in resolution by a factor of two have not been found to have
any affect upon our results.

Because we are concerned with energy transfer and dissipation, the form of
the artificial viscosity used in the models might be expected to play a
significant role.  In order to examine that possibility, we have computed
versions of Model~1
using both scalar and tensor forms of the artificial viscosity, with
the coefficient varied by a factor of two, and both with and without 
linear artificial viscosity.  The energy changes in the cloud and background
were found to differ among the models by no more than 10\%.  We therefore
conclude that the form of the artificial viscosity does not dominate our
results.  The lack of sensitivity is most likely to to the absence of
strong shocks in the models.

The models are run with reflecting boundary conditions on all sides.  This
choice of boundaries serves to isolate the system, eliminating
potential ambiguities in the interpretation of the energies of the two
components.

\section{Results of the Numerical Simulations}

\subsection{Model 1: Transsonic sedimentation of adiabatic clouds}

In Model~1, we adopt a polytropic equation of state for both the cloud and
background.  For the values of
$D_\rho$, $S$, and $g$ of the model, $V_t \sim C_h$ during the descent.

\begin{figure}[p]
\begin{center}
\includegraphics[width=6.0in]{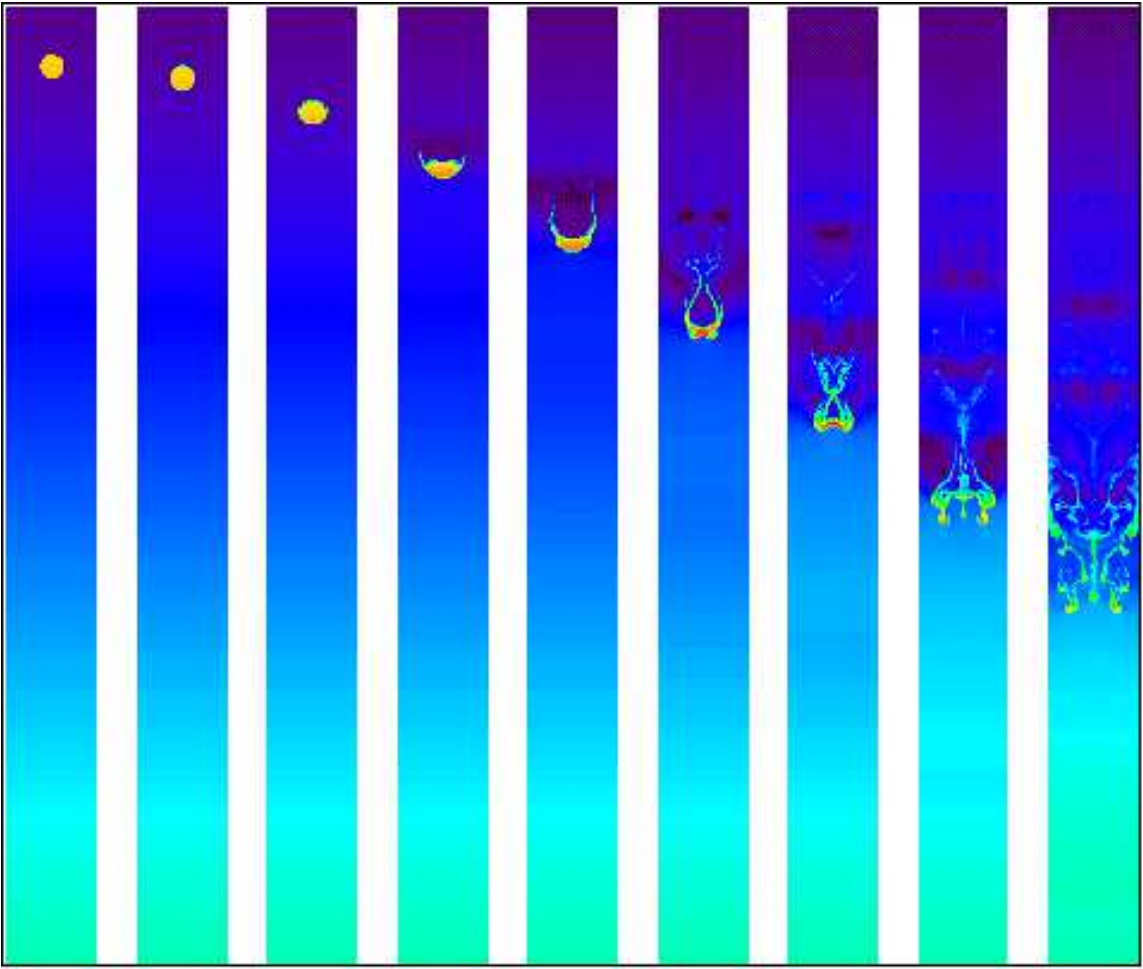}
\caption{Density evolution of Model~1.  The model is shown from t = 0
to 16, in intervals of 2, where the horizontal sound crossing time
is 1.5.}
\label{fig:mod1rho}
\end{center}
\end{figure}

In Figure~\ref{fig:mod1rho}, we show the evolution of the density of Model~1.
The model is shown from time 0 to 16, at intervals of 2 (the horizontal
sound crossing time in the RHG $\Delta x/C_h=1.5$).  The WFC rapidly
accelerates to a speed $\vert V_y\vert \approx C_h$, at which point
the increasing drag causes it to achieve a terminal speed.  The deceleration
of the cloud as it approaches terminal speed leads to the growth of
Rayleigh-Taylor instability, causing rapid breakup of the
cloud.  For an incompressible fluid, the Rayleigh-Taylor instability
grows, in the linear regime, as ${\rm e}^{\omega t}$, where
\begin{equation}
\omega^2={{2\pi g}\over \lambda}\left({{\rho_h-\rho_l}
\over{\rho_h+\rho_l}}\right),
\label{eq:rt}
\end{equation}
$\lambda$ is the wavelength of the perturbation, and the subscripts
$h$ and $l$ refer, respectively, to the heavy and light fluids
(Chandrasekhar 1961, p. 428).  For subsonic flows, the growth rate is
similar for compressible fluids.  Perturbations with the shortest wavelengths
grow most rapidly, but saturate quickly when their amplitudes
$A\approx\lambda$.  As a result, wavelengths $\lambda\sim S$
lead most strongly to cloud breakup.  For such perturbations, the
above relation gives $\omega\approx17$, in fair agreement with the
rate of breakup observed in the cloud, though the latter is complicated by 
the additional growth of Kelvin-Helmholtz instability due to the flow of
gas around the cloud (cf. Murray et al. 1993).

\begin{figure}[p]
\begin{center}
\includegraphics[width=6.0in]{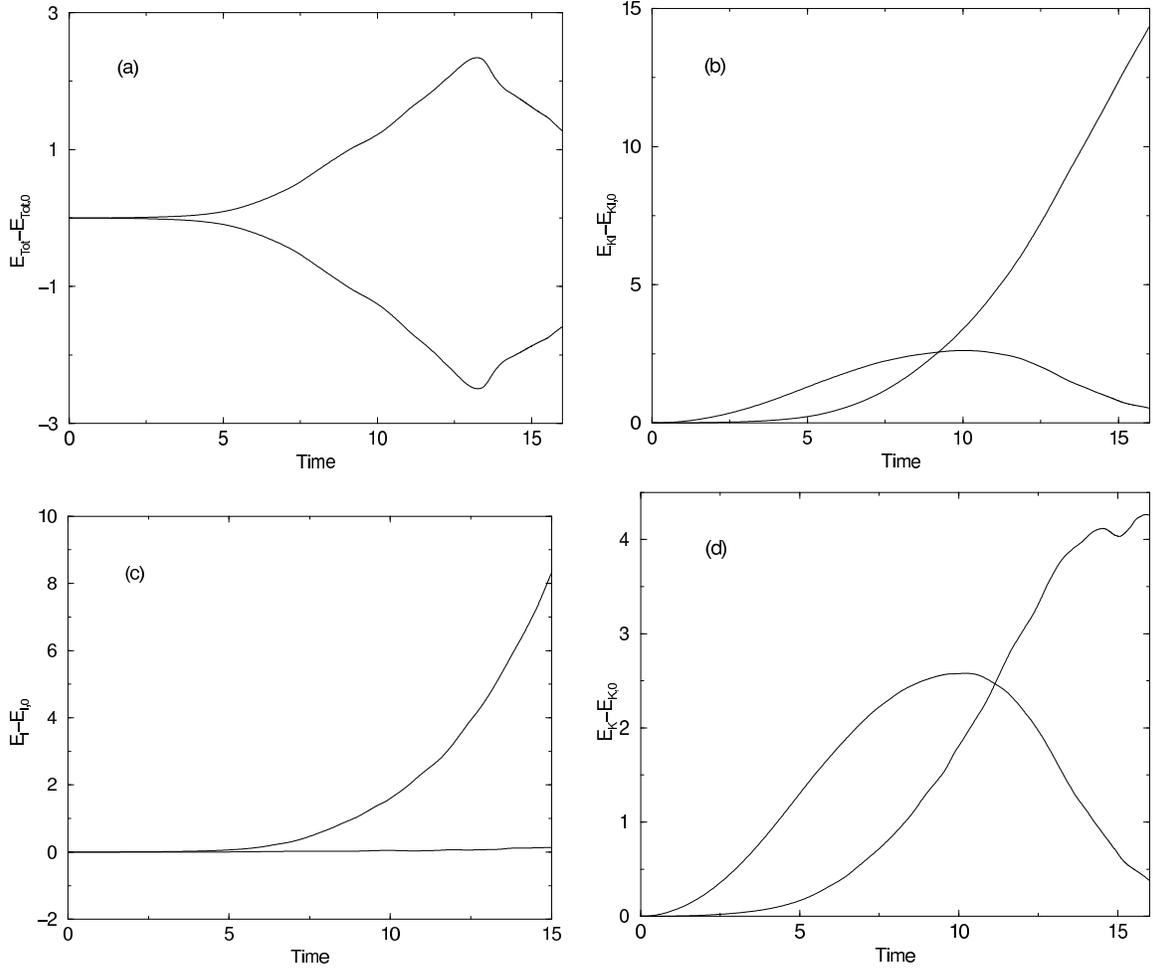}
\caption{Energy evolution of Model~1.  Shown are the time evolution of the
(a) total, (b) kinetic plus internal, (c) internal, and (d) kinetic
energies.  Data for the background gas are shown as the solid curves,
while that for the dense cloud are shown as dashed curves.}
\label{fig:mod1en}
\end{center}
\end{figure}

Figure~\ref{fig:mod1en} shows the energy evolution of Model~1.  Shown are
the evolution of the total (internal, kinetic, and gravitational), the
kinetic plus internal, kinetic, and 
internal energies.  Values for the background gas are given by the solid
curves, while those for the cloud are indicated by the dashed curves.  The
energies are in code units, and are plotted as changes relative to their
initial values.  The energies of the cloud and background are calculated
as sums across the entire computational grid, with the contribution from 
each zone weighted by the fractional amount of the appropriate tracer
present in each zone.  This should minimize any confusion due to mixing of
the cloud and background gas.  The high order of the advection scheme
also minimizes numerical diffusion (Anninos, Fragile, \& Murray 2003).

As can be seen in Figure~\ref{fig:mod1en}, the total energy of the
cloud decreases as it falls in the gravitational potential.  The
increase in $E_{Tot}$ at late times is due to the upward motion of
cloud material entrained within the vortices that form behind the
cloud.  The kinetic energy of the cloud increases until it reaches a
terminal infall speed at $t\approx10$, According to
Equation~(\ref{eq:vterm}), the terminal speed of infalling clouds is an
increasing function of their size.  As a cloud breaks up into
many smaller pieces, its kinetic energy decreases along with $V_t$.  The
internal energy of the cloud does not change significantly during its
descent and breakup. The distance travelled before breakup is $\sim 30
S(y_0)$.  The effective cross section of the cloud is $\sim 2 S(y_0)$,
implying that the mass of the RHG that is encountered by the
falling cloud is smaller than, but comparable to the mass of the cloud
(Murray et al. 1993).  During break up, the terminal velocity of
the fragments $V_t\propto S^{1/2}$, in accordance with
Equation~(\ref{eq:vterm}).  The fragments therefore trail behind the
remaining clouds.

The kinetic, and especially the internal energy of the background gas
are substantially increased by the end of the simulation.  In this
model, therefore, the majority of the energy released by the infall of
the cloud is deposited into the internal energy of the background gas,
primarily by the action of weak shock waves generated by the motion of
the infalling cloud.  This result supports the assumption of \cite{LM00}
that the rate of energy deposition throughout the galaxy is directly
proportional to the total infall rate of WFC's throughout the system.

\subsection{Model 2: Supersonic impact of WFC's}

In Model 1, the WFC attains a terminal speed which is a significant
fraction of both $C_h$, and the value of $V_t$ predicted from
Equation~\ref{eq:vterm}.  It might also be expected that WFC's which
condense from the RHG would have initial speeds comparable to the sound
speed of the RHG.  In order to examine the possible effects of nonzero
initial speeds upon the evolution, we consider in Model~2 an
initial condition in which the WFC is already falling at the sound
speed at the start of the numerical calculation.

\begin{figure}[p]
\begin{center}
\includegraphics[width=6.0in]{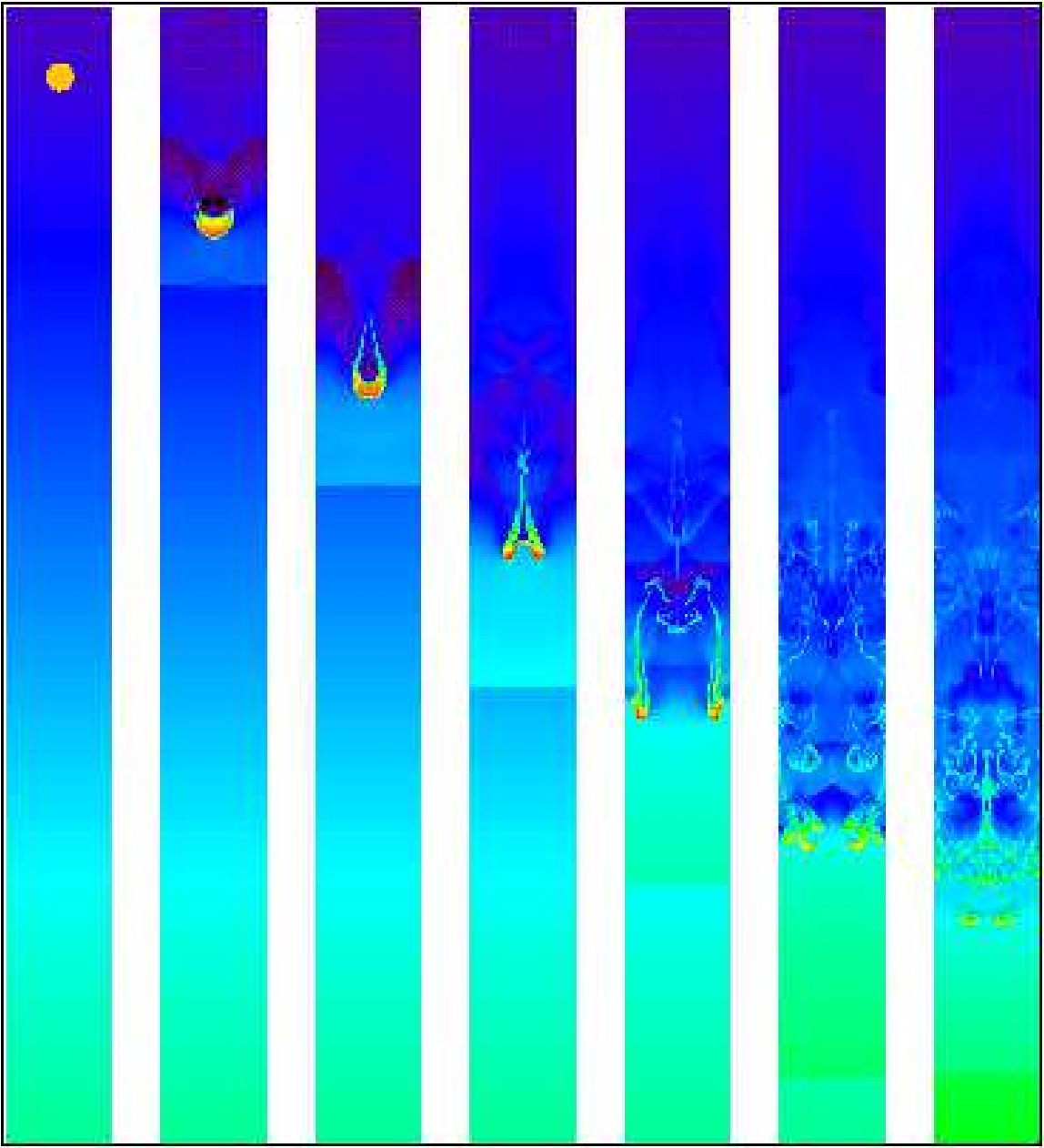}
\caption{Density evolution of Model~2, displayed as in
Figure~\ref{fig:mod1rho}.  Due to the rapid motion of the
cloud, the simulation is only carried out to $t=12$.}
\label{fig:mod2rho}
\end{center}
\end{figure}

The density of Model~2 is shown in Figure~\ref{fig:mod2rho}.  Due to
the more rapid motion of the cloud, as compared to that of Model~1,
the simulation is only carried out to $t=12$.  The initial motion of
the cloud can be seen to drive a weak shock ahead of it.  Behind the
shock, the leading edge of the cloud continues to move downwards at
almost $C_h$, slowing down gradually until the very end of the
simulation, when it rapidly decelerates as it breaks up, due to the
combined action of Rayleigh-Taylor and Kelvin-Helmholtz instabilities.
These results suggest that the infalling WFC's quickly settle to $V_t$
irrespective of the initial conditions, as we have assumed previously
(Lin \& Murray 2000).  The breakup of the cloud proceeds at nearly the
same vertical height as in Model~1.  The similarity arises because the
models have the same gravitational accelerations and density contrast.
Prior to breakup,
the downward motion of the cloud is more rapid than the value of $V_t$
found for Model~1.  The differences are due to the modification in the
drag caused by the leading shock in Model~2.

The energy evolution is shown in Figure~\ref{fig:mod2en}.  The initial
kinetic energy of the cloud, $E_{K,0}=6.3$, is almost entirely
dissipated by $t=10$.  Over the same time interval, the cloud is also
able to penetrate to a greater depth than the cloud in Model~1,
increasing the release of gravitational energy relative to that model.
Together, these effects lead to a gain of internal energy for the
background gas by a approximately a factor of two larger than seen in
Model~1.

However, the depth at which the cloud breaks up is similar in the two
models.  As in Model~1, the break up occurs when the cloud encounters a 
column of RHG that is comparable in mass to that of the
cloud.  Thereafter, the fragments' rate of sedimentation is
significantly reduced in accordance with Equation~(\ref{eq:vterm}).  The
asymptotic rate of RHG's internal energy increase in Model 2 is
comparable to that in Model 1.  

\subsection{Model 3: Efficient energy loss within the cool clouds}

\begin{figure}[p]
\begin{center}
\includegraphics[width=6.0in]{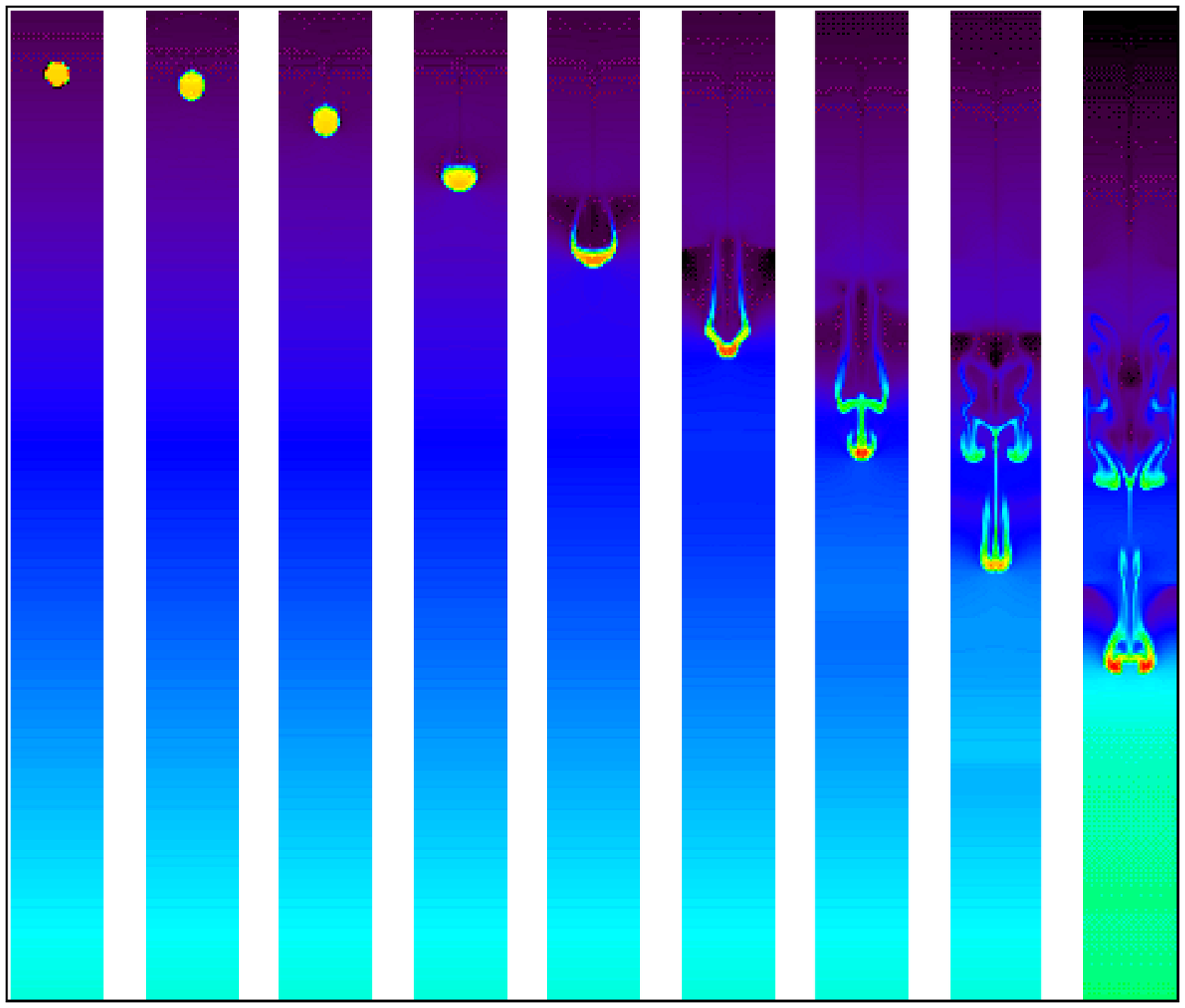}
\caption{Density evolution of Model~3, displayed as in
Figure~\ref{fig:mod1rho}.}
\label{fig:mod3rho}
\end{center}
\end{figure}

In Model~3, we approximate an isothermal equation of state for the cloud, as
described above, in order to represent the limit in which cooling is highly
efficient.  The evolution of the density is shown in Figure~\ref{fig:mod3rho},
while the energies are shown in Figure~\ref{fig:mod3en}.  The isothermal
behavior of the cloud leads to nonconservation of the total energy of the
cloud plus background, and so we do not plot that here, focusing instead
upon the kinetic and internal energies.

As expected, cooling within mixed cells does lead to some cooling of
the background gas, as well as some overcooling within the cloud, both
of which can be seen in Figure~\ref{fig:mod3en}.  The lack of heating
within the cloud leads to additional compression relative to the
previous models, reducing its breakup.  Overall, however, the transfer
of kinetic energy of the cloud to the internal energy of the
background gas is very similar to the adiabatic models described
above, indicating that efficient cooling within the clouds does not
have a strong effect upon the energy deposition rate.

Fragmentation of the cloud also occurs in Model 3. The efficient
cooling enhances the density contrast between the cloud and the RHG,
such that the cloud retains smaller volume and cross section.  Consequently,
the cloud encounters a smaller gas mass along the path of its descent,
and fragmentation occurs at a greater depth.  On small wavelengths, the
infalling cloud appears to be better preserved than in the previous models.
But on the scale of the cloud size, the cloud again fragments after
encountering a column similar to its own mass, as above.

\subsection{Model 4: Subsonic Sedimentation of WFC's}

\begin{figure}[p]
\begin{center}
\includegraphics[width=6.0in]{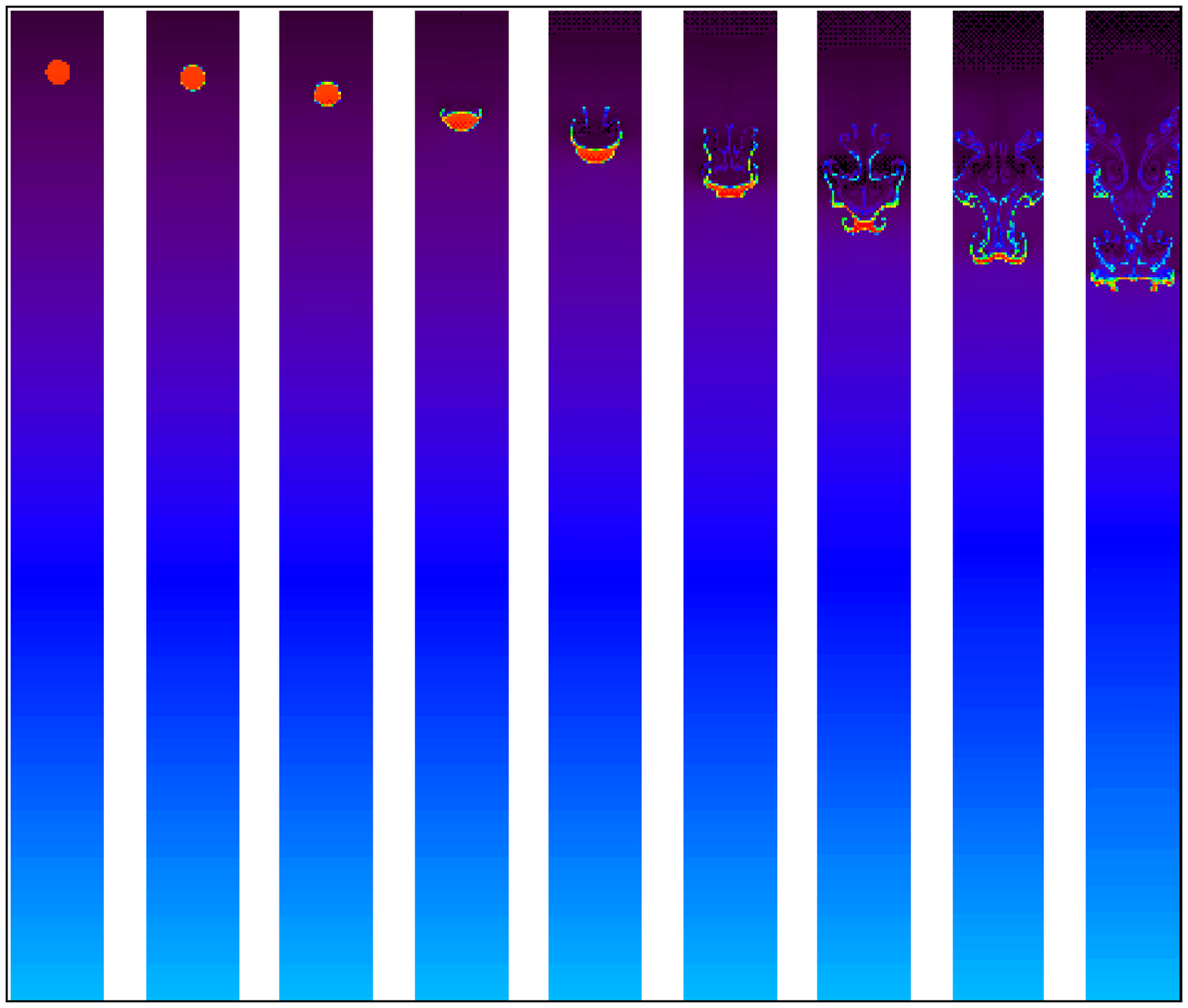}
\caption{Density evolution of Model~4, displayed as in
Figure~\ref{fig:mod1rho}.}
\label{fig:mod4rho}
\end{center}
\end{figure}

For Model~4, $D_\rho=25$, and $g=0.05$, such that $V_t$ is predicted by
Equation~\ref{eq:vterm} to be
subsonic.  The evolution of the density of Model~4 is shown in
Figure~\ref{fig:mod4rho}.  The cloud
rapidly reaches a terminal speed, $V_t\approx0.3$, smaller than
predicted if $C_D=1$.  As in Model~1, however, expansion of the cloud
enhances the drag coefficient to $C_D>1$.  The cloud therefore never
achieves the terminal speed predicted for
a hard sphere, even in the absence of strong supersonic dissipation.
From Equation~\ref{eq:rt}, $\omega\approx6$, and the cloud
breaks up even more rapidly than the more dense clouds considered in
Models~1-3, due to its reduced density contrast relative to those models.

The downward displacement of the cloud in Model~4 is reduced by a factor of a
few relative to that of Model~1.  As a result,
the gravitational energy released by the settling of the cloud, and 
dissipated into the background gas, is reduced by an order of magnitude
relative to Model~1, as can be seen in Figure~\ref{fig:mod4en}.

In the absence of trans/supersonic motion of the cloud through the
background, shock dissipation cannot be a strong mechanism for the
dissipation of energy due to motion of the cloud.  The primary mechanism
involves the wake of the cloud.  In the simulations, the vortical motions
behind the cloud dissipate energy on small scales, due to artificial
and numerical viscosity.  In three dimensions, the high Reynolds numbers
would lead to the formation of turbulent wakes, which would lead to the
dissipation of energy by viscous stress on sufficiently small length 
scales, leading to the same outcome as observed in Model~4.  The observed
outcome of the energy deposition is not, therefore, sensitive to the
exact physical process responsible for it.

\section{Summary and Discussion}

In this paper, we examine the interactions of a two-phase medium in a
passive gravitational potential.  This situation represents the physical
environment that occurs naturally in the context of galaxy formation,
cooling flows, and during the transition of gas clouds from quasi-hydrostatic
contraction to dynamical collapse.  It is a natural consequence of thermal
instability, which generally leads to the emergence of a population of
relatively cool, dense clouds (warm, fragmentary clouds, or WFC's) that are
pressure confined by an ambient hot, tenuous gas (residual hot gas, or RHG).
In such a state, the hot gas establishes a 
quasistatic equilibrium with the background gravitational potential, and
the cold clouds settle into it under the action of their negative buoyancy.  
In the present investigation, we neglect the self-gravity of the gas, and
consider the potential to be due to a time-invariant background distribution
of dark matter or stars.

Through a series of numerical simulations, we demonstrate the following
evolutionary outcomes. 

\noindent
1) During their descent, the WFC's break up on the same timescale 
as is required for them to attain a terminal speed.

\noindent
2) Most of the energy released from the sedimentation of the WFC's 
into the background gravitational potential is deposited into the RHG.

These results provide justifications for the assumptions we made in
our earlier model for the evolution of multi-phase medium during the
epoch of galaxy formation (Lin \& Murray 2000).  They also resolve an
outstanding conceptual issue with regard to the energy source needed for
the persistence of the multi-phase medium.  In particular, the RHG can
achieve a thermal equilibrium, in which its heat loss via bremsstrahlung
emission and conduction into the WFC's is balanced by the release of
energy from the infalling WFC's.  This equilibrium allows
a multi-phase structure to be maintained in the system.  

The equilibrium is very dynamic.  The WFC's are continually formed by
thermal instability within the hot gas.  As they move within the hot gas,
they break up, and are eventually re-heated by conduction from the hot
gas.  They may also cool and form stars, if local sources of UV radiation
are lost.  Additionally, the fragmentation of the WFC's increases their
surface area to volume ratio.  This reduces their timescale for 
collisions and mergers, leading to
the formation of larger WFC's.  A natural extension of the present
investigation, therefore, is to consider the collisional equilibrium for a
population of WFC's.  We shall investigate this in future work.

\begin{acknowledgements}
This work was performed under the auspices of the U.S. Department of
Energy by University of California, Lawrence Livermore National
Laboratory under Contract W-7405-Eng-48.  This work is partially
supported by NASA through an astrophysical theory grant NAG5-12151.
\end{acknowledgements}

\clearpage

\begin{deluxetable}{crrrc}
\tablewidth{0pt}
\tablecaption{Model Parameters\label{tab:models}}
\tablehead{
\colhead{Model} &
\colhead{$g$} &
\colhead{$\gamma_w$} &
\colhead{$D_\rho$} &
\colhead{$V_{y,0}/C_h$} 
}
\startdata
1 & 0.10 & $5\over3$ & 100 & {\phantom -}0 \\
2 & 0.10 & $5\over3$ & 100 & -1 \\
3 & 0.10 &  1        & 100 & {\phantom -}0 \\
4 & 0.05 & $5\over3$ &  25 & {\phantom -}0 \\
\enddata
\end{deluxetable}

\clearpage

\clearpage

\clearpage

\clearpage

\begin{figure}[p]
\begin{center}
\includegraphics[width=6.0in]{figure4.ps}
\caption{Energy evolution of Model~2, displayed as in 
Figure~\ref{fig:mod1en}.}
\label{fig:mod2en}
\end{center}
\end{figure}

\clearpage

\clearpage

\begin{figure}[p]
\begin{center}
\includegraphics[width=6.0in]{figure6.ps}
\caption{Energy evolution of Model~3, displayed as in 
Figure~\ref{fig:mod1en}.}
\label{fig:mod3en}
\end{center}
\end{figure}

\clearpage

\clearpage

\begin{figure}[p]
\begin{center}
\includegraphics[width=6.0in]{figure8.ps}
\caption{Energy evolution of Model~4, displayed as in
Figure~\ref{fig:mod1en}.}
\label{fig:mod4en}
\end{center}
\end{figure}

\end{document}